\documentclass{pasa}%
\usepackage{lineno}

\title[PKS 1954$-$388]{PKS 1954$-$388: RadioAstron Detection on 80,000\,km Baselines and Multiwavelength Observations}
\author[Edwards et al.]{P.G.~Edwards$^1$, 
  Y.Y.~ Kovalev$^2$,
  R.~Ojha$^{3,4,5}$,
  H.~An$^{6,19}$,
  H.~Bignall$^1$,
  B.~Carpenter$^{3,4}$,
  T.~Hovatta$^{7,8}$,
  J.~Stevens$^1$,
  P.~Voytsik$^2$,
  A.S.~Andrianov$^2$, 
  M.~Dutka$^{3,4}$,
  H.~Hase$^9$,
  S.~Horiuchi$^{10}$,
  D.L.~Jauncey$^{1,11}$,
  M.~Kadler$^{12}$,
  M.~Lisakov$^2$,
  J.E.J.~Lovell$^{13}$,
  J.~McCallum$^{13}$,
  C.~M{\"u}ller$^{14}$,
  C.~Phillips$^1$,
  C.~Pl{\"o}tz$^{12}$,
  J.~Quick$^{15}$, 
  C.~Reynolds$^1$,
  R.~Schulz$^{16}$,
  K.V.~Sokolovsky$^{2,17,18}$,
  A.K.~Tzioumis$^1$
  \and
  V.~Zuga$^2$
  \\
\affil{$^1$ CSIRO Astronomy and Space Science, PO Box 76, Epping, NSW 1710, Australia}%
\affil{$^2$ Astro Space Center of Lebedev Physical Institute, Profsoyuznaya 84/32, 117997 Moscow, Russia}
\affil{$^3$ NASA, Goddard Space Flight Center, Greenbelt, MD, 20771, USA}
\affil{$^4$ Catholic University of America, Washington, DC, 20064, USA}
\affil{$^5$ University of Maryland, Baltimore County, 1000 Hilltop Cir, Baltimore, MD 21250, USA}
\affil{$^6$ Kavli Institute for Particle Astrophysics and Cosmology, Stanford University, Stanford, CA 94305, USA}
\affil{$^7$ Aalto University Mets{\"a}hovi Radio Observatory, Mets{\"a}hovintie 114, FL-02540 Kylm{\"a}l{\"a}, Finland}
\affil{$^8$ Aalto University Department of Radio Science and Engineering, PO Box 13000, FI-00076 Aalto, Finland}
\affil{$^9$ Bundesamt f{\"u}r Kartographie und Geod{\"a}sie,  93444, Bad K{\"o}tzting, Germany}
\affil{$^{10}$ CSIRO Astronomy and Space Science, Canberra Deep Space Communication Complex,  ACT 2901, Australia}
\affil{$^{11}$ Research School of Astronomy and Astrophysics, Australian National University, Canberra ACT, 2611, Australia}
\affil{$^{12}$ Lehrstuhl f{\"u}r Astronomie, Universit{\"a}t W{\"u}rzburg, 97074, W{\"u}rzburg, Germany}
\affil{$^{13}$ School of Physical Sciences, University of Tasmania, Private Bag 37, Hobart, TAS 7001, Australia}
\affil{$^{14}$ Department of Astrophysics/MAPP, Radboud University Nijmegen, PO Box 9010, 6500 GL, Nijmegen, The Netherlands}
\affil{$^{15}$ Hartebeesthoek Radio Astronomy Observatory, Krugersdorp 1740, South Africa}
\affil{$^{16}$ ASTRON, the Netherlands Institute for Radio Astronomy, PO Box 2, 7990 AA Dwingeloo, Netherlands}
\affil{$^{17}$ IAASARS, National Observatory of Athens, 15236 Penteli, Greece}
\affil{$^{18}$ Sternberg Astronomical Institute, Moscow State University, 
  119992 Moscow, Russia}
\affil{$^{19}$ Department of Astronomy and Space Science, Chungbuk National University, Cheongju 28644, Republic of Korea}
}%
\jid{PASA}
\doi{10.1017/pas.\the\year.xxx}
\jyear{\the\year}

\begin{document}%

%
\begin{abstract}
We present results from a multiwavelength study of the blazar PKS
1954$-$388 at radio, UV, X-ray, and gamma-ray energies.  A RadioAstron
observation at 1.66\,GHz in June 2012 resulted in the detection of
interferometric fringes on baselines of 6.2 Earth-diameters.  This
suggests a source frame brightness temperature of greater than
2$\times$10$^{12}$\,K, well in excess of both equipartition and
inverse Compton limits and implying the existence of Doppler boosting
in the core.  An 8.4\,GHz TANAMI VLBI image, made less than a month
after the RadioAstron observations, is consistent with a previously
reported superluminal motion for a jet component.  Flux density
monitoring with the Australia Telescope Compact Array confirms
previous evidence for long-term variability that increases with
observing frequency.  A search for more rapid variability revealed no
evidence for significant day-scale flux density variation.  The ATCA
light-curve reveals a strong radio flare beginning in late 2013 which
peaks higher, and earlier, at higher frequencies.  Comparison with the
{\it Fermi} gamma-ray light-curve indicates this followed $\sim$9
months after the start of a prolonged gamma-ray high-state --- a radio
lag comparable to that seen in other blazars. The multiwavelength data
are combined to derive a Spectral Energy Distribution, which is fitted
by a one-zone synchrotron-self-Compton (SSC) model with the addition
of external Compton (EC) emission.
\end{abstract}
\maketitle%
\section{INTRODUCTION }
\label{sec:intro}

A major challenge in astronomy is the struggle to observe objects with
an angular resolution sufficient to probe the underlying physical
mechanisms. The longer wavelengths of radio-astronomy initially made
the quest for high angular resolution more difficult, but the relative
ease of preserving phase information enabled the technique of Very
Long Baseline Interferometry (VLBI). Intercontinental VLBI routinely
achieves milli-arcsecond--scale angular resolutions, and extending the
baselines between telescopes into space, with satellite-based
telescopes, currently yields the highest angular resolution achieved
in astronomy.

In this paper we describe multi-wavelength studies of PKS 1954$-$388
with the Australia Telescope Compact Array ({\bf ATCA}, maximum baseline 6\,km),
and present a second-epoch 8.4\,GHz TANAMI
{\bf (Tracking Active Galactic Nuclei with Austral Milliarcsecond Interferometry)}
VLBI image (maximum
baseline $\sim$10,000\,km) and a 1.66\,GHz RadioAstron Active Galactic
Nuclei (AGN) Survey observation (maximum baseline $\sim$80,000\,km, or
6.2 Earth-diameters).  The pronounced variability of the source at
radio and gamma-ray energies is examined, and we combine these with
data from the {\it Swift} and {\it Fermi} satellites to obtain a
Spectral Energy Distribution (SED) for the source which is fitted by a
self-consistent model.

\section{PKS 1954$-$388 }
\label{sec:1954}

PKS 1954$-$388 was first catalogued in the Parkes 2.7\,GHz survey
\cite{shim71}. The source was observed in May 1969 with a flux density
of 2.00$\pm$0.05\,Jy and again in June 1970 with a flux density of
1.50$\pm$0.05\,Jy, and noted as being possibly variable. A comparison
with the Molonglo 408\,MHz catalog indicated an inverted spectrum,
prompting follow-up optical observations with the Mt Stromlo 74\,inch
(1.9\,m) telescope yielding an identification with an 18th magnitude
galaxy \cite{sbpw71}. Early observations with the 3.9\,m
Anglo-Australian Telescope gave a redshift of $z$=0.63 \cite{bsb75}.

The discovery of a high level of optical polarisation, 11\%, by Impey
\& Tapia \shortcite{it88} led to classification of the source as a
blazar.  Oshlack, Webster, \& Whiting \shortcite{oww02} used
measurements of the H$\beta$ line width and luminosity to estimate a
central black hole mass of 4.3$\times$10$^8$M$_\odot$ for the source.

PKS 1954$-$388 was not detected by the EGRET instrument on the {\it
  Compton Gamma-Ray Observatory}. However, the radio properties led
Vercellone et al. \shortcite{ver04} to classify the source as a
candidate gamma-ray AGN, a prediction borne out by the detection of
the source with the {\it Fermi} Large Area Telescope \cite{1fgl}.
Studies by Nolan et al. \shortcite{nol12} indicate the source is
variable at gamma-ray energies, with a $<$1\% chance of the flux being
steady, behaviour common to the AGN class.

The source was included in the VSOP {\bf (VLBI Space Observatory Programme)}
AGN Survey \cite{hir00} though it
was not observed before the end of {\bf the} mission.  Inclusion
in the survey list did, however, result in the source being part of
the multi-epoch monitoring program with the ATCA at 1.4, 2.5, 4.8 and
8.6\,GHz between 1996 October and 2000 February. This revealed
pronounced variability at the higher frequencies but no evidence at
1.4\,GHz for any variability on time scales of $\sim$70 days
\cite{tin03}.

A series of VLBI observations ---
at 4.9\,GHz in 1993 May \cite{shen98},
at 4.9\,GHz in 1996 June \cite{fom00},
at 15\,GHz in 1998 June \cite{kov05},
and at  2.3 \& 8.6\,GHz in 2002 December \cite{push12} ---
found the source to be strongly core dominated
(see also Pusharev \& Kovalev 2015).
A deeper VLBI observation at 8.4\,GHz in 2002 July detected,
in addition to the core, a marginally significant
second component $\sim$2.8\,mas to the west 
\cite{ojha04}.

Piner et al.\ \shortcite{pin12} used 36 epochs of Radio Reference
Frame VLBI observations at 8\,GHz between 1994 and 2003 to measure
apparent speeds for two jet components.  Both components displayed
motions of $\sim$0.1\,mas yr$^{-1}$, with corresponding apparent
speeds of 3.7$\pm$1.8\,$c$ and 3.7$\pm$1.0\,$c$ for the fainter, outer
and brighter, inner components respectively.  The outer component of
Piner et al.\ \shortcite{pin12} is consistent with the weak secondary
component of Ojha et al.\ \shortcite{ojha04}.

PKS 1958$-$388 is part of the multi-epoch TANAMI program and a first
epoch 8.4\,GHz image from 2008 February is presented by Ojha et
al.\ \shortcite{ojha10}. A jet extends over $\sim$5\,mas to the west,
and a source frame brightness temperature of 1.5$\times$10$^{12}$\,K
was derived for the core.  B{\"o}ck et al.\ \shortcite{boe16} report a
source frame brightness temperature of 2.2$\times$10$^{11}$\,K for a
subsequent TANAMI observation in 2008 November, indicating significant
variability.  The brightness temperature is the surface brightness of
a radio source expressed for convenience in terms of the equivalent
black body temperature. It is an important parameter as there are
limits on the intrinsic brightness temperature imposed by both
Inverse-Compton cooling and equipartition arguments --- see, e.g.,
Kellermann \shortcite{kel02} for a review.  As the measureable maximum
brightness temperature depends on baseline length, the best way to
constrain measurements is to extend VLBI baselines beyond the Earth's
surface with an element in space.  The RadioAstron mission achieved
this with the launch of the Spektr-R satellite on 2011 July 18, with
observations at 0.3, 1.6, 4.8 and 22\,GHz being routinely conducted
\cite{kar13}.

\section{OBSERVATIONS}

\subsection{RadioAstron}

A RadioAstron AGN Survey Program observation of PKS 1954$-$388 was
made at 1.66\,GHz on 2012 August 23 with observation code raes03br.
The snap-shot mode observation started at 15:45\,UT and lasted one
hour using the Spektr-R satellite\footnote{
The 1.6\,GHz receiver on the satellite was designed by
the CSIRO Division of Radiophysics with the support of CSIRO Office of
Space Science and Applications and manufactured by British Aerospace
Australia, with the low-noise amplifier built by Mitec
Ltd.\ Australia.}
and the Parkes 64\,m and Mopra 22\,m telescopes.  
At 1.66\,GHz, the satellite formats RCP data with
16-MHz bandwidths, which are 1-bit sampled. A parity bit is added for
each byte and the data transmitted to the Earth at 144\,Mbps.  The
ground telescopes record data in two, dual-polarisation, 16-MHz
bandwidth IF bands, with 2-bit sampling of the data yielding a data
rate of 256\,Mbps.

Data from the spacecraft were transmitted in real time to the
Pushchino tracking station and recorded to disk. Data from the ground
radio telescopes were recorded to disk and later reformatted to Mark
5B format before being electronically transferred to the Astro Space
Centre (ASC).  The data were correlated at the ASC correlator
\cite{kar13}, with clear fringes being found at 1.6\,GHz between the
three telescopes (see Figure~1).  The data were exported from the
correlator and fringe-fit in {\bf the software package} PIMA \cite{pima}.  The ($u,v$)-coverage
for the observation is, as expected for this snap-shot mode, very
sparse, with the projected Parkes--Mopra baseline being
$\sim$1\,M$\lambda$ and the baselines to the spacecraft ranging from
403 to 440\,M$\lambda$, 5.8 to 6.2 Earth-diameters, at a position
angle of $\sim$177$^\circ$.  The goal of the AGN Survey is to
determine whether sources are detectable over a range of baseline
lengths and use this information to infer angular sizes and
corresponding brightness temperatures.

\begin{figure}
\begin{center}
\includegraphics[width=\columnwidth,angle=0]{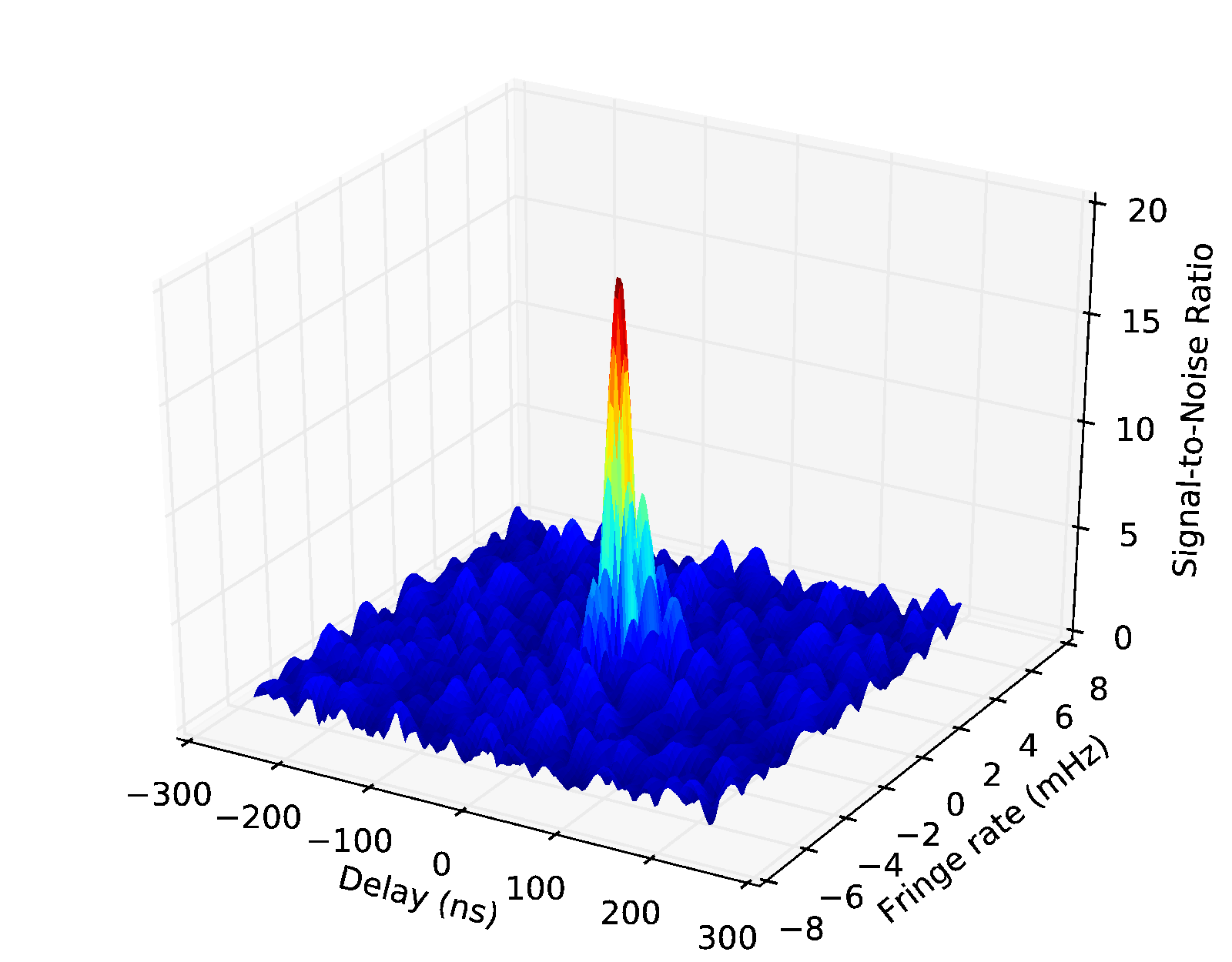}
\caption{The detection of fringes at 1.66\,GHz in (delay, delay-rate) space for PKS\,1954$-$388
  between the Parkes 64\,m telescope and the RadioAstron satellite, Spektr-R.
}\label{Fig1}
\end{center}
\end{figure}

\begin{figure}
\begin{center}
\includegraphics[angle=0,width=\columnwidth]{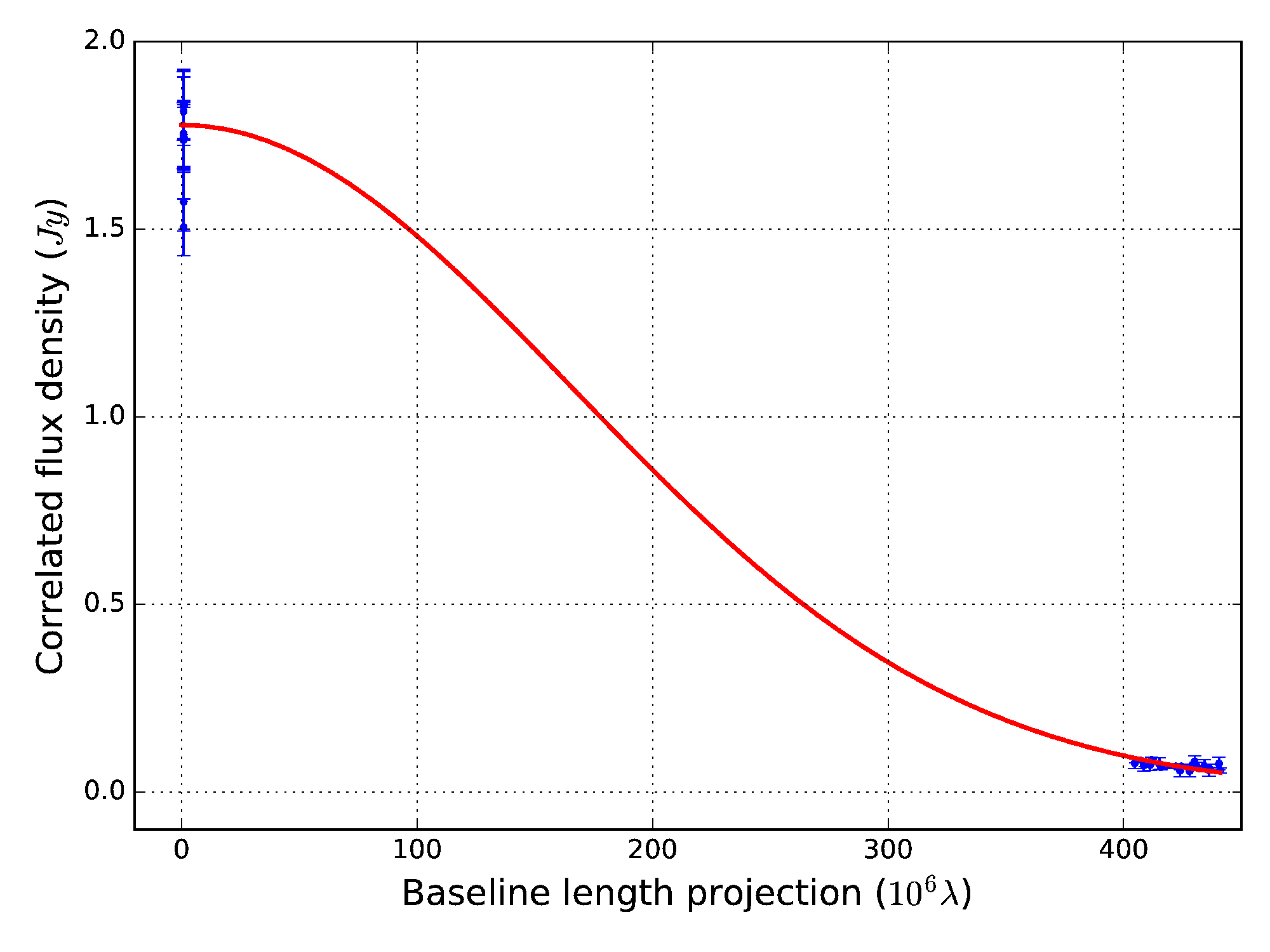}
\caption{Plot of the correlated flux density as a function of baseline
  length for PKS\,1954$-$388 at 1.66\,GHz.  Blue points are the data:
  the inner points correspond to the Parkes--Mopra baseline, the outer
  points are the baselines to the RadioAstron satellite, Spektr-R The
  red curve is the simplest Gaussian model-fit --- see text for
  details.  }
 \label{Fig2}
\end{center}
\end{figure}

Initially the amplitudes and signal-to-noise ratios (SNRs) were
plotted as a function of integration time for each baseline on each
scan in order to estimate the coherence time. For this observation, no
significant losses were evident with up to 10 minutes integration
time.  Space VLBI fringes were detected in 10 minute integration times
with SNRs of $\sim$20 on the baselines to Parkes, and SNRs of $\sim$7
on the baselines to Mopra, corresponding to a probability of false
detection on SVLBI baselines significantly less than 10$^{-6}$.

A correlated flux density of 1.78\,Jy was measured on the
Parkes--Mopra baseline, and $\sim$0.07\,Jy on Earth--space baselines.
The simplest model-fit to the data is with single circular Gaussian
component with a total flux density of 1.78\,Jy and full-width
half-maximum of 0.47\,mas (see Figure~2).  A circular Gaussian
component was adopted for simplicity, although the size is tightly
constrained in the north-south direction but poorly determined in the
east-west direction.  Following Kovalev et al.\ \shortcite{kov05}, we
calculate an observer's frame brightness temperature of
3.6$\times$10$^{12}$\,K for this model.  The limited ($u,v$)-coverage
means we cannot be certain that fitting a single Gaussian
model-component to the ground-ground and ground-space data is the
correct approximation, and in fact it is highly likely that the
north-south Parkes--Mopra baseline is sampling some of the jet
structure to the west of the core (see Sections~2 and 3.2).  However,
there is a very conservative minimum brightness temperature associated
just with the ground-space data: following Lobanov \shortcite{lob15},
we calculate a formal minimum observer's frame brightness temperature
of 1.3$\times$10$^{12}$\,K.  These values must be multiplied by
(1+$z$) to obtain the corresponding source frame brightness
temperatures.  Incorporating the expected calibration uncertainties,
the simple fit, which effectively provides an upper limit, yields a
source frame brightness temperature of (6$\pm$1)$\times$10$^{12}$\,K
and the minimum source frame brightness temperature is
2$\times$10$^{12}$\,K respectively.

\subsection{TANAMI}

An 8.4\,GHz TANAMI observation was made on 2012 September 16, less
than a month after the RadioAstron observation. Participating in this
observation were the ``tied array'' of five 22\,m ATCA antennas, a
single 12\,m ASKAP
{\bf (Australian Square Kilometre Array Pathfinder)}
antenna with a single pixel feed, the Ceduna 30\,m,
Hartebeesthoek 26\,m, Hobart 26\,m, Parkes 64\,m, Tidbinbilla 70\,m,
TIGO 6\,m telescopes, and the 12\,m Katherine telescope of the AuScope
array \cite{lov13}.

PKS\,1954$-$388 was observed in six $\sim 10$ minute snapshots over a
period of 9 hours to build up good $(u,v)$-coverage. The source had
not risen at Hartebeesthoek (in South Africa) for the first three
scans and had set at the TIGO site (in Concepcion, Chile) for the last
three scans. The resulting image is shown in Figure~3.

\begin{figure}
\begin{center}
\includegraphics[width=\columnwidth]{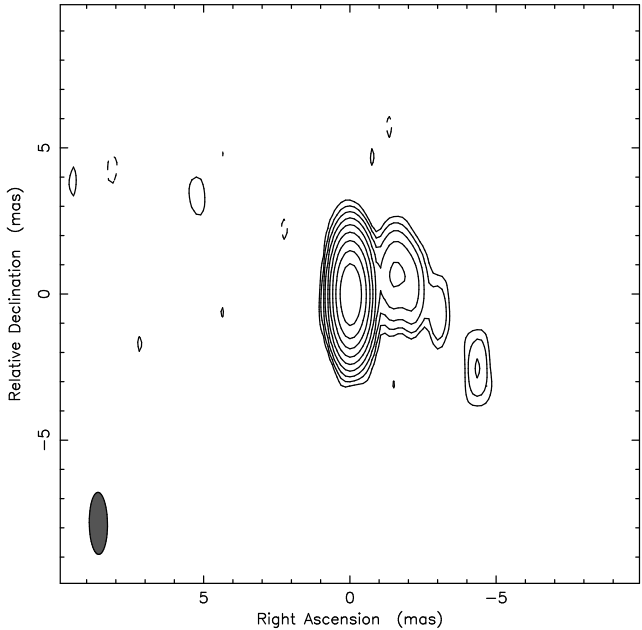}
\caption{TANAMI image from {\bf an 8.4\,GHz} observation on 2012 September 16.  The
  image peak is 1.29\,Jy/beam, and the beam (FWHM) is 2.1\,mas
  $\times$ 0.6\,mas at a position angle of 1$^\circ$.  Contour levels
  are $-$0.2 (dashed), 0.2, 0.4, 0.8, 1.6, 3.2, 6.4, 12.8, 25.6 and
  51.2\% of the peak.  }
 \label{Fig3}
\end{center}
\end{figure}

Model-fitting of this TANAMI observation yields a 1.5\,Jy core, an
85\,mJy component at 1.7\,mas from the core at a position angle of
$-$67$^\circ$, and a 42\,mJy component at 2.1\,mas and a position
angle of $-$101$^\circ$. There is also about 11\,mJy of flux 5\,mas
from the core and a position angle of $-$120$^\circ$. This morphology
is consistent with previous VLBI images in Ojha et
al.\ \shortcite{ojha04}, Ojha et al.\ \shortcite{ojha10} and Piner et
al.\ \shortcite{pin12} all of which report a bright core and a
westward jet.

The innermost component is readily identifiable with the innermost
component of Piner et al.\ (labelled 2 in their Table 5) given the
similarity in intensity, position angle, and extrapolated core
distance.  The other components do not lie at the extrapolated
distance ($\sim$3.2\,mas) of the outer Piner et al.\ component.  Given
the sparse ($u,v$)-coverage of the Radio Reference Frame observations
for sources as far south as PKS 1954$-$388, the complex jet
morphology, and the large gap in observing epochs, the details of
proper motion are an open question to be resolved by further TANAMI
monitoring.

\subsection{ATCA}

PKS 1954$-$388 has been regularly observed with the ATCA as part of
the flux density calibration monitoring program (C007) and also as
part of a multi-frequency program (C1730) monitoring TANAMI and other
gamma-ray sources \cite{ste12}. Prior to the Compact Array Broadband
Backend (CABB: Wilson et al. \shortcite{wil11}) upgrade in 2009, C007
observations were made with 128\,MHz bandwidths centred at 4.8 \& 8.6,
18.5 \& 19.5\,GHz. In the CABB era, C1730 observations are made with
2\,GHz bandwidths centred at 5.5 \& 9, 17 \& 19, and 38 \& 40\,GHz.
Observations at 38 \& 40\,GHz commenced in early 2007.  The
observations are made in snap-shot mode, with integrations of several
minutes in each pair of bands. Observations at 38 \& 40\,GHz are
preceded by a ``pointing'' scan on a bright source (on occasions, PKS
1954$-$388 itself) to update the pointing model for that area of
sky. PKS 1934$-$638 is used as primary flux density calibrator
\cite{par16}, with Uranus also used in the 7mm band in more compact
array configurations.  Errors are dominated by systematic effects and
are estimated to be less than 5\% below 10\,GHz, and up to 10\% at
40\,GHz.  The data presented in this paper are incorporated in the
ATCA calibrator
database\footnote{http://www.narrabri.atnf.csiro.au/calibrators/},
which provides a flux density model of each source in each observed
band for each epoch. This allows us to compute the flux density at the
same frequency regardless of the centre frequency that was actually
observed in a particular epoch.

The resulting light-curve in four frequency bands is shown in
Figure~4.  The source is obviously quite variable and in a high-state
in 2005--2006.  From 2007 to 2013 the source showed some irregular
variations from epoch to epoch, which is undersampled by our
observations, but overlaid on a general decline in flux density.  We
note that the 2012 August RadioAstron observation was made when the
source was in a relatively quiescent radio state.  At the end of 2013
the source underwent another outburst.  In this instance the source
was monitored in all three frequency bands and, as shown in Figure~5
(an expansion of a sub-range of Figure~4), the radio flare proceeds
earlier, and more rapidly at higher frequencies.

As mentioned in Section~2, PKS 1954$-$388 was monitored at four
frequencies over 16 epochs by Tingay et al.\ \shortcite{tin03}.  The
source was in a high state ($\sim$5\,Jy at 8.4\,GHz, with an inverted
spectrum) in October 1996 however was at its lowest levels
($\sim$1\,Jy at all frequencies) in February 2000.  Tingay et
al.\ \shortcite{tin03} characterised the variability of sources
monitored at multiple epochs with a variability index defined as the
RMS variation from the mean flux density, divided by the mean flux
density.  PKS 1954$-$388 had variability indices of 0.11 at 1.4\,GHz,
0.16 at 2.5\,GHz, 0.26 at 4.8\,GHz, and 0.30 at 8.6\,GHz.  The
variability index at 8.4\,GHz was in the top 10\% of the sample of 185
sources.

We undertake a similar analysis for the ATCA data presented here.  In
order to make a comparison between frequencies, we only use the data
points after the installation of the 7mm receivers in 2007, and as a
consequence omit the data from the large flare in 2005--2006.

PKS 1954$-$388 had variability indices of 
0.13 at 5.5\,GHz (29 data points),
0.12 at 9.0\,GHz (29 data points),
0.16 at 17\,GHz (25 data points), and
0.19 at 38\,GHz (20 data points).
While the absolute values obviously differ
the general trend of the variability index increasing with frequency persists.

\begin{figure}
\begin{center}
\includegraphics[width=\columnwidth]{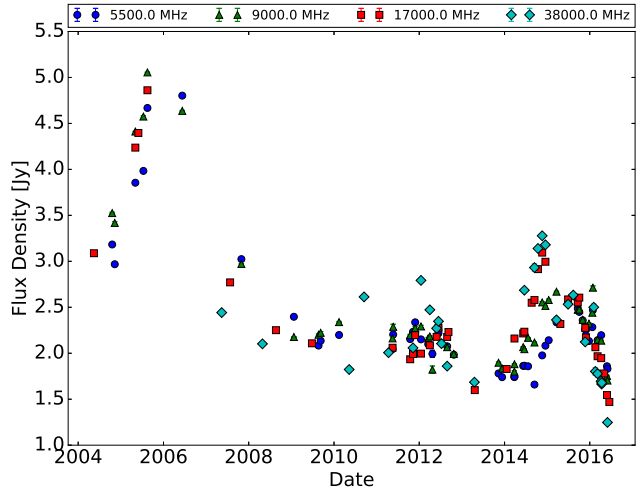}
\caption{The radio light-curve derived from ATCA monitoring data --- see text for details.
}
 \label{Fig4}
\end{center}
\end{figure}

\begin{figure}
\begin{center}
\includegraphics[width=\columnwidth]{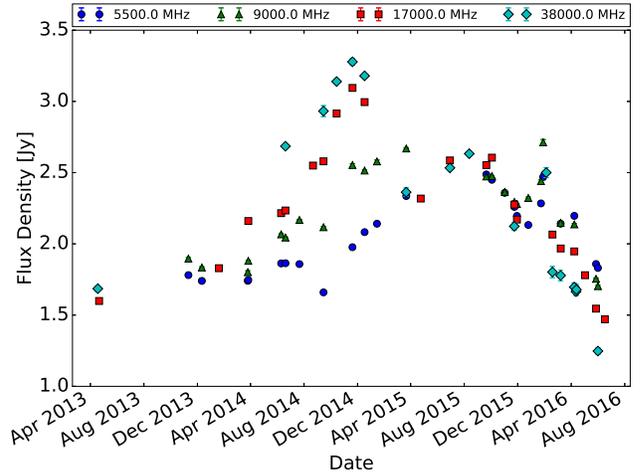}
\caption{A subset of the ATCA monitoring data in Figure~4 with an expanded
  time axis to more clearly see the frequency dependence of the flare
  that commenced at the end of 2013.
}
 \label{Fig5}
\end{center}
\end{figure}

\subsubsection{Search for short timescale variability with the ATCA}

PKS 1954$-$388 was also observed with the ATCA as part of project
C2898 (``Monitoring of RadioAstron AGN Survey targets to measure
intra-day and longer term variability'').  Short timescale variability
at a range of GHz frequencies potentially provides information on the
scattering properties of the local Galactic interstellar medium along
the line of sight.  PKS 1954$-$388 was observed in approximately 10
scans at each of several frequencies between 1 and 20\,GHz, with
irregular sampling over a 15-day period in 2014 June. It was also
observed as part of the same project several times between 2014 July
and September.

For the purposes of searching for short time-scale variability, since
we are not limited by signal-to-noise, a section of each frequency
band was selected that appeared to be relatively free of interference
and other problems.  Only the upper part of the 1--3\,GHz band was
selected for analysis, due to the lower frequencies being heavily
affected by radio-frequency interference in the data from June 2014;
the ATCA was in a compact configuration {\bf (EW352)} at this time.  The band
selection was identical for all epochs. Bandpass, amplitude gain and
polarization leakage corrections were determined for each day of
observations using several minutes of data on the ATCA primary
calibrator, PKS 1934$-$638. Since PKS 1954$-$388 is a bright, isolated
point source at the ATCA, the flux density was estimated by averaging
the visibilities over all baselines after applying phase-only
self-calibration.

Table~\ref{tab:IDV} shows central frequency, bandwidth used, mean flux
density $\bar{S}$ and RMS of the flux density measurements in June,
averaged over 1-minute intervals and over all channels in the selected
frequency range, and the variability index as defined previously,
$\mu = RMS/\bar{S}$.

\begin{table}[h]
\caption{Results of search for short-term variation with the ATCA over 15 days in 2014 June.}
\begin{center}
\begin{tabular}{@{}ccccc@{}}
\hline\hline
$\nu$    & Bandwidth & $\bar{S}$  & $S_{\sc RMS}$  & $\mu$ \\
(GHz)    & (MHz)     &  (Jy)      &    (Jy)        &  \\
\hline
2.75    &    500     &  1.613     &  0.025         & 0.016 \\
4.8     &    400     &  1.832     &  0.022         & 0.012 \\
8.8     &   1000     &  2.034     &  0.022         & 0.011 \\
17.0    &   2000     &  2.295     &  0.031         & 0.014 \\
19.0    &   2000     &  2.307     &  0.029         & 0.013 \\
\hline\hline
\end{tabular}
\end{center}
\label{tab:IDV}
\end{table}


During the ATCA observations in 2014 June, PKS~1954$-$388 showed RMS
variations no larger than 1.6\%, or 30\,mJy, on timescales between
minutes and days in the frequency range 2.5--20~GHz. These variations
are not significantly larger than those determined for other sources
observed in 2014 June, or than the systematic uncertainties determined
by comparing data on different baselines.  We assume the observed
variations here represent upper limits on the true variability.  A
significant, $\sim$5--10\%, decrease in flux density was observed over
a 10-day period in 2014 July, over a broad bandwidth between 2 and
20\,GHz, with the largest change around 5\,GHz. Such a change could be
consistent with interstellar scintillation of a compact component, but
with only two independent data points in this time period it is not
possible to constrain the variability mechanism. The spectrum became
significantly more inverted in 2014 September, with a drop in flux
density at frequencies below 7\,GHz and an increase at frequencies
above 8\,GHz. This period is associated with a flare at higher
frequencies seen in the longer-term ATCA monitoring data (see
Figure~5).

\subsection{Swift}

We have analysed archival \textit{Swift} \cite{geh04} data in order to
better characterise the SED of PKS\,1954$-$388. The X-Ray Telescope
(XRT; Burrows et al.\ 2004) on board the \textit{Swift} spacecraft is
a grazing incidence telescope with an effective area of 110 cm$^{2}$,
a field of view of 23.6 arcmin and 0.2--10\,keV energy range. It
observed PKS\,1954$-$388 on 2007 July 31 for 5.911 ksec.

The UltraViolet and Optical Telescope (UVOT; Roming et al.\ 2004) is a
30\,cm UV/optical telescope co-aligned with the XRT covering the
170--650 nm range in a 17$'$ x 17$'$ field. The UVOT observed
PKS\,1954$-$388 on 2010 April 9 for 1.184 ksec.

The XRT data were processed using the \texttt{xrtpipeline} and the
spectrum was fit using \texttt{XSpec}, both a part of the
\texttt{HEASoft} package (v.6.12). A circular region with a radius of
20 pixels was used to extract the source events while a background
region with a 50 pixel radius accounted for background events. The
spectrum had $>$200 counts and so was fitted using chi-squared
statistics.  The spectrum was fit in the 0.3--10 keV region using an
absorbed power law. The N$_{H}$ column density was set to $6.43 \times
10^{20}$\,cm$^{-2}$ \cite{kalberla2005}.  The ancillary response files
were generated with \texttt{xrtmkarf}.

The UVOT data were reduced using the \texttt{uvotsource} task, also a
part of the \texttt{HEASoft} package (v.6.12). Source counts were
extracted from a circular region with a 10 pixel radius, the
background was extracted from a circular region with a 30 pixel
radius.

\subsection{Fermi}

The gamma-ray data were obtained with the Large Area Telescope (LAT)
aboard {\it Fermi}, which observes the entire sky every 3 hours at
energies of $0.03-300$\,GeV \cite{atwood09}. The publicly available
Pass 8 data\footnote{http://fermi.gsfc.nasa.gov/ssc/data/} were
analysed using the {\it Fermi} ScienceTools software package version
v10r0p5. We used the instrument response functions P8R2\_SOURCE\_V6,
Galactic diffuse emission model ``gll\_iem\_v06.fits'' and isotropic
background
model\footnote{http://fermi.gsfc.nasa.gov/ssc/data/access/lat/BackgroundModels.html}
``iso\_P8R2\_SOURCE\_V6\_v06.txt''.  Following the LAT data selection
recommendations\footnote{http://fermi.gsfc.nasa.gov/ssc/data/analysis/documentation
  /Cicerone/Cicerone\_Data\_Exploration/Data\_preparation.html}, we
select photons in the event class 128 and use a zenith angle cut of
90$^\circ$.

Photon fluxes in the $0.1-300$\,GeV energy range were calculated using
unbinned likelihood analysis and the tool ``gtlike'' with the Minuit
optimizer. We used a 7-day binning in the light-curves and calculated
a 2$\sigma$ upper limit if the test statistic (TS) value was less than
4 (e.g., Abdo et al.\ 2011). The source model was generated with the
tool\footnote{https://fermi.gsfc.nasa.gov/ssc/data/analysis/user/}
``make3FGLxml.py'' with all sources within 20$^\circ$ of
PKS\,1954$-$388 included in the model with their spectral parameters,
except the flux, frozen to the values determined in the 3rd {\it
  Fermi} LAT catalog (3FGL; Acero et al.\ 2015). For sources more than
10$^\circ$ from PKS\,1954$-$388 we also froze the fluxes to the 3FGL
values.

\begin{figure*}
\begin{center}
\includegraphics[width=15cm]{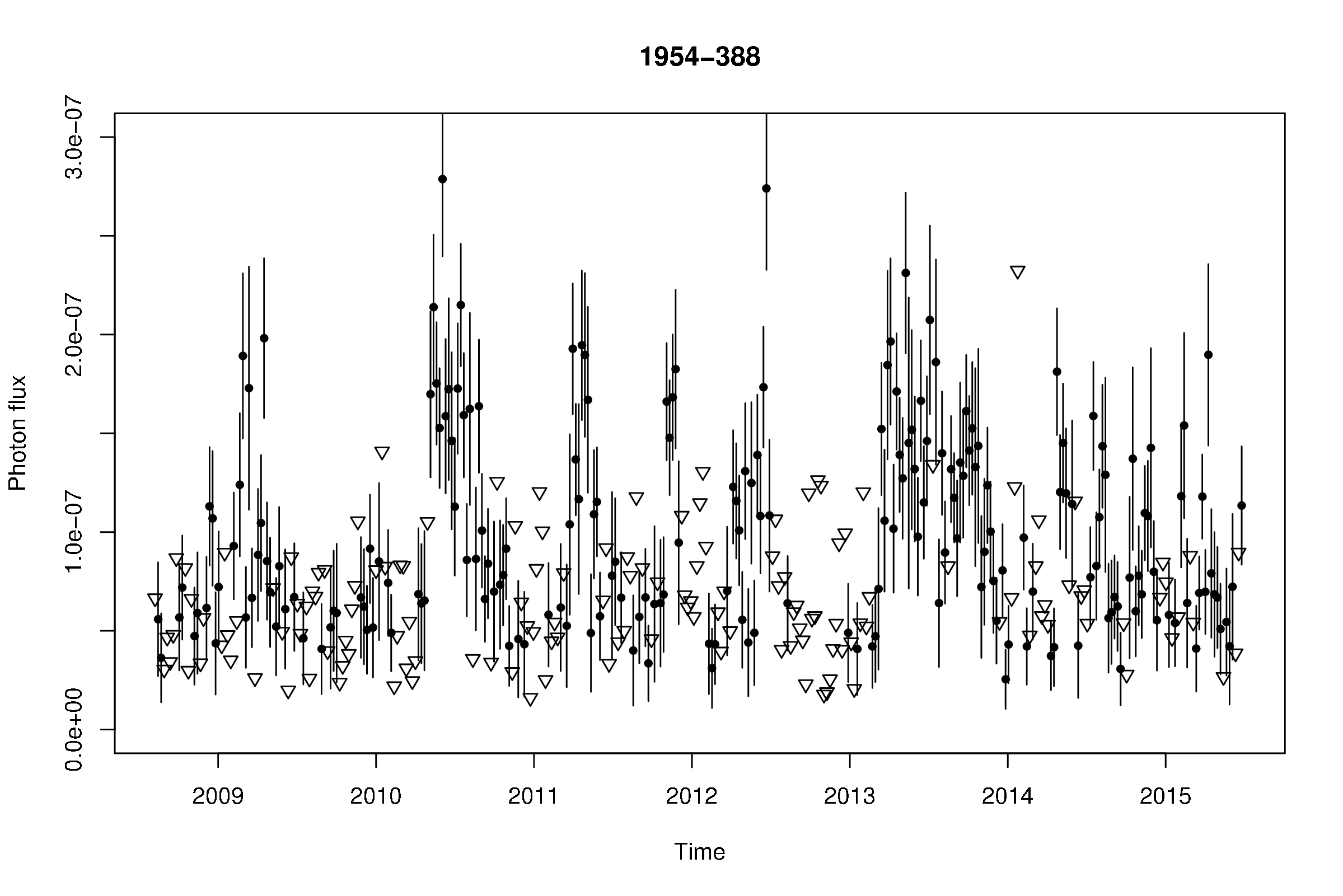}
\caption{{\it Fermi} light-curve with one-week bin width.  Photon
  fluxes are plotted in units of cm$^{-2}$\,s$^{-1}$.  Inverted
  triangles denote 2$\sigma$ upper limits --- see text for details.  }
 \label{Fig6}
\end{center}
\end{figure*}

The resulting light-curve is shown in Figure~6.  The data in the
figure span 2008 August 8 to 2015 June 30.  There are 360 week-long
bins over this period.  Significant fluxes were measured for 210
weeks, with upper limits derived for the remaining 150 weeks.

Until early 2013, the photon flux can be characterised as
having short periods of enhanced flux with longer 
periods of quiescent flux. There is 
a $\sim$4 week flare in 2009 March, 
a $\sim$12 week flare starting in 2010 May, 
a $\sim$9 week flare in 2011 April/May, 
a $\sim$4 week flare in 2011 November, and   
a $\sim$12 week flare starting in 2012 April.
It is notable that the onsets for the second and fourth flares
are more rapid than revealed by the 7-day bin-width.
There are several periods containing a series of upper limits
--- late 2009, early 2010, the beginning of 2011,
beginning of 2012 and late 2012 --- during which the 
average weekly photon flux drops below a significant detection.

The behaviour from 2013 March onwards has been quite different.  An
outburst (again with a rapid onset) started in mid-March and persisted
until the end of that year. For the 18 months since then the state has
changed much more frequently.  It is tempting to postulate that this
changed behaviour is an effect of the prolonged 2013 high state.  We
discuss this further in Section~4.3.

\subsection{SED}

We combine the data from the previous sections with other multiwavelength data
to derive the broadband SED
and model it using one-zone synchrotron-self-Compton (SSC)
emission of an
electron/positron component with an addition of external Compton (EC)
emission \cite{boe97}. In this model, the e$^{-}$/e$^{+}$ component is
injected and propagates along the jet axis for a specified time
period. While propagating, the particles lose their energy via
synchrotron radiation and inverse-Compton upscattering. The seed
photons for the inverse-Compton upscattering are the synchrotron
radiation of the electrons themselves (SSC), and the emission of the
Shakura-Sunyaev disk \cite{sha73} and the broad-line region (EC). We
follow the component for $10^{7}$\,s ($l=3\times10^{17}$\,cm),
calculate the integrated spectrum, and match the resulting SED with
the observed one.  We show our models and the parameters in Figure~7
and Table~2.  Note that the data are not acquired contemporaneously.
The {\it Fermi} data used here are the average fluxes for the period
from 2008 August 4 to 2015 October 9.

The model parameters are not well constrained because the parameters
are covariant.  Hence, a different set of parameters may also be able
to describe the observed SED reasonably.  Furthermore, the SED is not
a contemporaneous one, and large variability of blazars prevents us
from inferring the parameters accurately. Nevertheless, we find that
the SED is well described with a typical electron injection spectrum
having a power-law index of $p\sim2$. In the model, the optical
($\sim10^{13}-10^{14}$\,Hz) and the X-ray spectra are the synchrotron
and the SSC component, respectively.  Then, simple extension of the
SSC component to higher energy explains the {\it Fermi} spectrum
reasonably (Figure~7 left), and the EC components do not play a
significant role in the {\it Fermi} LAT band, being distinguishable
only in the MeV band.  Because the EC emission of the disk photons has
a different shape from the observed high-energy SED, we need to
suppress this by using relatively large magnetic field ($B$), a small
bulk Lorentz factor ($\Gamma$) and placing the base of the jet far
from the black hole ($h_{\rm inj}$).  The radio emission is assumed to
be from a separate self-absorbed zone; hence the model does not
explain the radio data. However, radio data may provide important
insights into understanding blazars via variability studies.

We also show an alternative model in Figure~7 (right), where we use
the EC of broad-line region (BLR) photons to explain the {\it Fermi}
SED. In this case, the jet base needs to be farther from the disk to
suppress the disk EC emission. The magnetic-field strength for this
model is lower, and the SSC emission at higher energies is suppressed.
The data fit both models to within the measurement and modelling
uncertainties, and thus we cannot reasonably prefer one model over the
other.  Obtaining a contemporaneous SED and/or monitoring spectral
variability may help to distinguish between the models.  This has
recently been achieved by Krau{\ss} et al.\ \shortcite{kra16}, who
derive an SED for the relatively quiescent period from late 2008 to
early 2010 and compare it with an SED from a more active period from
mid-2010 to mid-2011, though without fitting a physical model to the
SEDs.

\begin{figure*}
\center
\begin{tabular}{cc}
\hspace*{-5 mm}
\includegraphics[width=9cm]{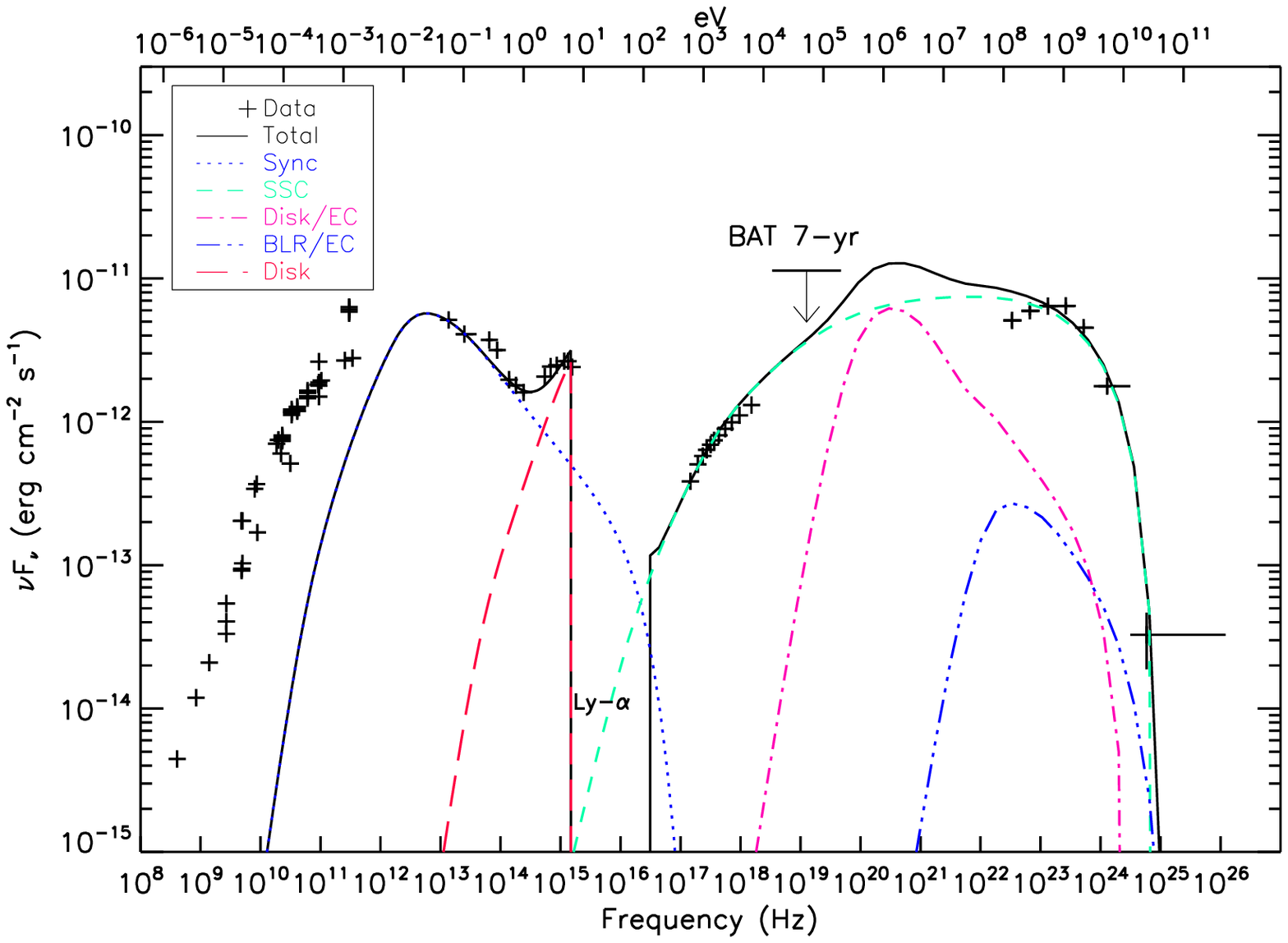} &
\hspace*{-7 mm}
\includegraphics[width=9cm]{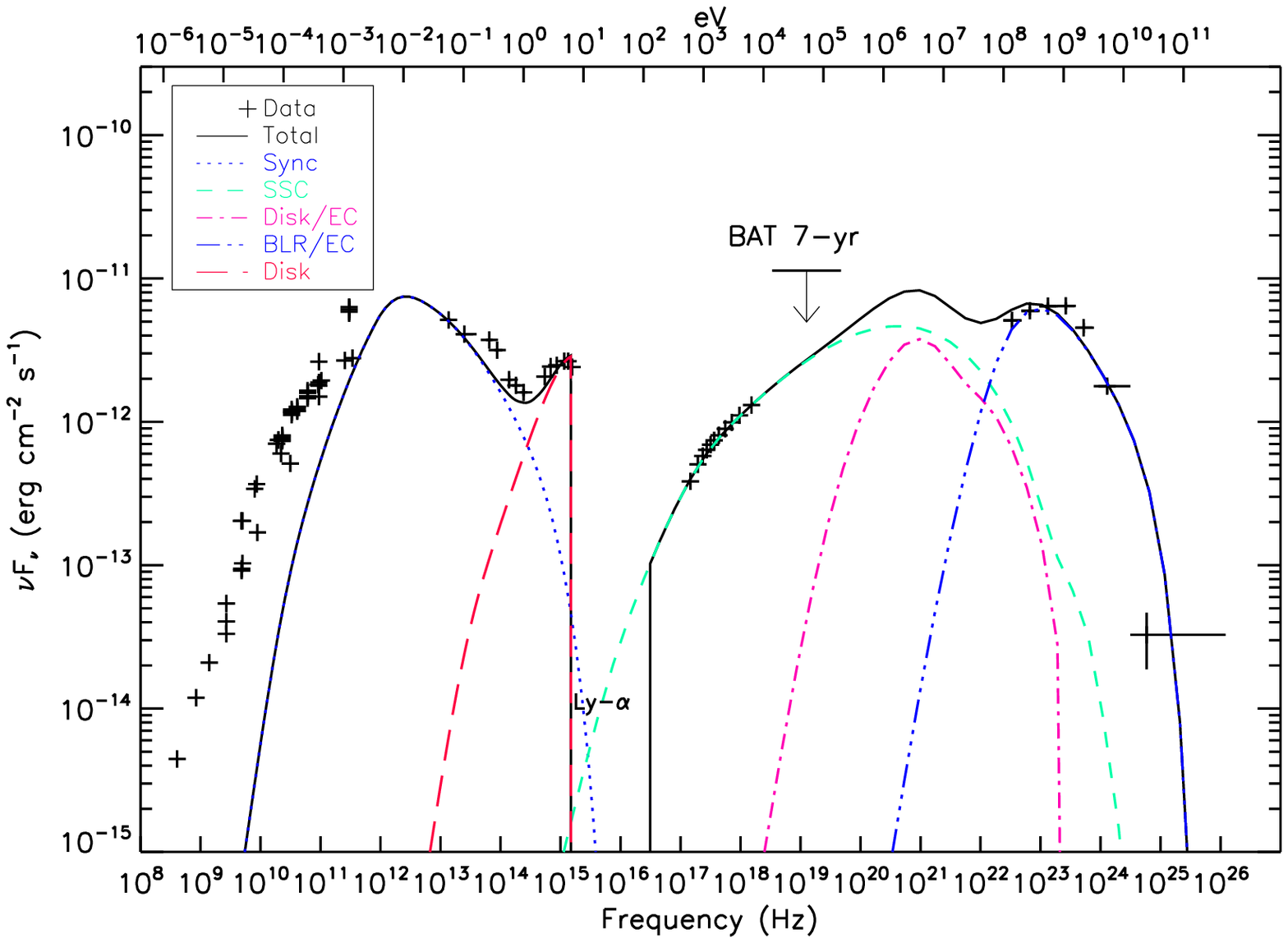} \\
\end{tabular}
\vspace{-3mm}
\caption{The observed SED and the models. The SED is measured in
  several radio bands, {\it Swift}/XRT, and {\it Fermi}. Infrared data
  are taken from the {\it WISE} and {\it 2MASS} catalogs. We also show
  the {\it Swift}/BAT 7-yr sensitivity.  Left: a model with the
  high-energy SED fit with SSC emission (model A). Right: a model with
  the high-energy SED fit with EC of BLR photons (model B). In these
  figures, the measured SED is plotted with crosses and the best-fit
  model is the black solid line. Individual model components are also
  shown: the blue dotted line is the synchrotron component, red dashed
  line is the direct disk component, pink dot-dashed line is the EC of
  the disk photons, blue triple-dot-dashed line is the EC of BLR
  photons, and the cyan dashed line is the SSC component.}
\label{Fig7}
\vspace{0mm}
\end{figure*}

\begin{table*}[t]
\vspace{-0.0in}
\begin{center}
\caption{Model parameters for the observed SED.
\label{tab:sed}}
\vspace{-0.05in}
\scriptsize{
\begin{tabular}{@{}lccc@{}} \hline\hline
Parameter                               & Symbol              & Model A            & Model B \\ \hline
Redshift                                & $z$                 & 0.63               & 0.63 \\
Bulk Lorentz Factor                     & $\Gamma$            & 3                  & 5 \\
Viewing angle (deg.)                    & $\theta_v$          & 5                  & 5 \\
Magnetic Field (G)                      & $B$                 & 1.5                & 0.4 \\
Comoving radius of component (cm)       & $R'_b$              & $6\times10^{14}$   & $2.6\times10^{14}$ \\
Initial Electron Spectral index         & $p_{\rm 1}$         & 2                  & 1.9 \\
Initial Minimum Electron Lorentz Factor & $\gamma'_{\rm min}$ & $1 \times 10^{3}$  & $1.4\times10^{3}$ \\
Initial Maximum Electron Lorentz Factor & $\gamma'_{\rm max}$ & $3 \times 10^{4}$  & $9\times10^{3}$  \\
Injection height (pc)                   & $h_{\rm inj}$       & 0.04               & 0.18 \\
Disk luminosity (erg s$^{-1}$)          & $L_{\rm disk}$      & $3 \times 10^{49}$ & $2.3\times 10^{49}$ \\
Mass of the black hole ($M\odot$)       & $M_{\rm BH}$        & $5 \times 10^{8}$  & $1.5\times10^{9}$ \\
Accretion efficiency                    & $\eta$              & 0.03               & 0.03 \\ \hline
\end{tabular}}
\end{center}
\hspace{-2.0 mm}
\end{table*}

\section{DISCUSSION}

\subsection{Brightness temperature and scattering}

The lower limit to the source frame brightness temperature of
2$\times$10$^{12}$\,K is well in excess of the equipartition
brightness temperature limit of $\sim$5$\times$10$^{10}$\,K
\cite{rea94} and the inverse Compton limit of $\sim10^{11.5}$\,K
\cite{kel69,rea94}, implying minimum Doppler factors of 40
(equipartition) and 6 (inverse Compton).  The Lorentz factor of 5 and
angle to the line of sight of 5$^\circ$ used in Model~B of the SED
fitting (Section~3.6) implies a Doppler factor of $\sim$9, and is
therefore consistent with the inverse Compton Doppler factor and with
the reported apparent speed of $\sim$3.7$c$ (Section~2).  The
equipartition Doppler factor is much more difficult to reconcile.

It is necessary though to consider the impact of scattering of the
source by the interstellar medium (ISM). The simple circular Gaussian
modelling of the RadioAstron observations of PKS 1954$-$388 used in
Section~3.1 yields a core size of 0.47\,mas, and if we assume that
only a fraction of the flux density on the Parkes--Mopra baseline (but
all of that on the space baselines) corresponds to the core, this
would result in an even more compact core.  The fitted angular size is
remarkably consistent with the angular broadening predicted by the
NE2001 model for Galactic electron density \cite{cor02} for the \lq
average' visibility regime \cite{goo89,tay93} for the line of sight to
PKS~1954$-$388 ($l=1.6^{\circ}, b=-29^{\circ}$). Although the ISM has
complex structure, and the NE2001 model cannot be used to reliably
determine scattering properties along a particular line of sight, the
result suggests that the intrinsic core size at 1.66\,GHz might be
significantly smaller than that measured, and the brightness
temperature correspondingly higher.

However, as discussed by Johnson \& Gwinn \shortcite{jon15}, it is
also necessary to consider the effect of refractive substructure
introduced by scattering in the ISM. Refractive substructure has been
seen in some RadioAstron pulsar observations \cite{gwi16,pop17} and
may also be present in the RadioAstron observations of 3C\,273
\cite{jon16}. This can have the effect of enhancing detections on long
baselines to RadioAstron, due to the appearance of fine-scale
structure in the image of a resolved, scattered source, leading to an
over-estimate of intrinsic brightness temperature.

We note that Pushkarev \& Kovalev \shortcite{push15} derive a median
core size for PKS 1954$-$388 of 1.0$\pm$0.2\,mas at 2.3\,GHz, and
0.35$\pm$0.05\,mas at 8.4\,GHz, and a size-frequency dependence of
$\nu^{-0.94\pm0.11}$, based on multi-epoch, simultaneous 2.3 and
8.4\,GHz observations. Their results imply that intrinsic source size
dominates over scatter-broadening at frequencies above 2.3\,GHz. The
upper limit to the angular size of 0.47\,mas inferred from the
RadioAstron observations of PKS 1954$-$388 is a factor of three times
smaller than the angular size expected at 1.66\,GHz from an
extrapolation of the Pushkarev \& Kovalev \shortcite{push15} fit. The
apparent discrepancy in core size measurement could be due to several
factors: (i) the size of the core component may be time variable, as
the flux density of the source is highly variable and was relatively
low during the RadioAstron observations; (ii) the core component
measured by Pushkarev \& Kovalev \shortcite{push15} resolves into
several components at sub-mas angular resolution, implying the
structure is more complex than accounted for by the simple circular
Gaussian model-fitting of the sparse data; or (iii) the RadioAstron
long baseline visibilities might be dominated by refractive noise. The
observed correlated flux density seen on the RadioAstron baselines is
at least $\sim$4\% of the flux density of the compact component, which
is considerably larger than the $\sim$1\% RMS refractive noise
predicted by Johnson \& Gwinn \shortcite{jon15}.  In order to
distinguish among the various possibilities, further RadioAstron
observations on intermediate baselines less than 6 Earth-diameters are
needed.

Although PKS 1954$-$388 is quite compact, has a flat radio spectrum
and is very variable, as is seen in the light-curves in Figures 4 and
5, the lack of any rapid, large-amplitude intra-day or inter-day
variability evident in ATCA monitoring in 2014 June (Section 3.3.1)
means there is no observational evidence for the intra-day variability
(IDV) often associated with scattering in the ISM.  It is not yet
possible to determine whether this is due to insufficiently compact
structure in the source, or the lack of a nearby region of enhanced
scattering along this line of sight, or both.

In any case, the detection of PKS 1954$-$388 on $\sim$80,000\,km
baselines at 1.6~GHz goes some way to vindicating the decision to
place the satellite in an eccentric 9-day orbit, with apogee heights
up to 350,000\,km. The long space baselines provided by RadioAstron
observations have enabled significantly higher brightness temperatures
to be measured, e.g., in excess of 10$^{13}$\,K for 3C\,273
\cite{kov16}, and the RadioAstron AGN Survey will extend these studies
to a much larger sample of sources (Kovalev et al., in
preparation). The high sensitivity of the 18\,cm receiver on board the
RadioAstron Space Radio Telescope, in combination with large
ground-based telescopes, is also leading to a new understanding of the
effects of interstellar scattering.

\subsection{Consideration of VLBI imaging}

VLBI imaging to date has revealed the source to be strongly core
dominated, with several weaker jet components. The multi-epoch
monitoring of Piner et al. \shortcite{pin12} found evidence for two
persistent components with apparent superluminal speeds. Extrapolating
these motions, of $\sim$0.1\,mas/yr, back to find an ejection epoch
yields 1992.3$\pm$2.4 for the inner component, and approximately 10
years earlier for the outer component \cite{pin12}.  These components
are thus unlikely to be associated with the bright radio flares in
1996 \cite{tin03} and 2005--2006 (this paper).  The extensive
multi-epoch parsec-scale monitoring of AGN by the MOJAVE
{\bf (Monitoring Of Jets in Active galactic nuclei with VLBA Experiments)}
program
demonstrated that, while there is some spread in the apparent speeds
of separate features within an individual jet, there is support for
the idea that there is a characteristic flow that describes each jet
\cite{lis09}.  Assuming, therefore, a similar motion for a component
that might have emerged from the core in 1996, it would have been
$\sim$1.2\,mas from the core at the time of the first epoch TANAMI
observation in 1998. It is interesting to note that there was indeed
evidence for a component $\sim$1\,mas from the core at that epoch.
The TANAMI image from 2012 presented in this paper does not clearly
resolve this component, which may be due to the slightly poorer
east-west angular resolution at that epoch, or to this component
merging with the inner of the two Piner et al.\ components.  Any
component ejected in 2005 would have been only 0.3\,mas from the core
in 2008 and 0.7\,mas from the core in 2012 and so barely
distinguishable from the core at these angular resolutions.

\subsection{Connection between gamma-ray and radio flares}

The short-term gamma-ray flares between 2009 and 2013 may have been
accompanied by similar short-duration radio flares; however our ATCA
monitoring has not been sensitive to such short events. The much
longer lived gamma-ray flare in 2013 is followed by a significant and
extended flaring at radio energies.  The characteristic brightening
earlier, and more rapidly at higher frequencies is consistent with
injection of a new population of high-energy electrons.  There is a
delay of approximately 9 months between the beginning of the gamma-ray
outburst and the start of the radio flare.

Max-Moerbeck et al.\ \shortcite{max14} cross-correlated 3-year {\it
  Fermi} light-curves with 4-year light-curves at 15\,GHz, finding
that only one of 41 sources with high-quality data in both bands shows
correlations with significance larger than 3$\sigma$, demonstrating
that great care is needed when comparing light curves even when many
years of data are used.  Thus, while it is a natural interpretation
that the gamma-ray and radio flares are causally related, it is not
possible to determine this unequivocally.

Fuhrmann et al.\ \shortcite{fuh14} conducted a cross-correlation of
radio and gamma-ray light-curves for 54 {\it Fermi} blazars, finding
that the average source-frame time delay between gamma-ray and radio
(with radio lagging) decreased from 76$\pm$23\,days at cm bands to
7$\pm$9\,days in mm/sub-mm bands, with this frequency dependence being
in good agreement with jet opacity dominated by synchrotron
self-absorption.  The ATCA data shown in Figure~5 follow this trend,
though the sampling is insufficiently frequent to assess this more
quantitatively.

Similarly, Ramakrishnan et al.\ \shortcite{ram15} examined the
discrete correlation function between radio and gamma-ray
light-curves, finding that in most sources, the gamma-ray peaks lead
the radio with time lags in the range 20 to 690 days. The
$\sim$9-month delay between the onset of the gamma-ray and radio
flares is thus quite consistent with these findings.

\section{SUMMARY}

The RadioAstron AGN Survey snapshot observation of PKS 1954$-$388 in
2012 detected the source on a 6.2 Earth-diameter baseline, confirming
the presence of a compact core.  A source frame brightness temperature
of greater than 2$\times$10$^{12}$\,K is implied, suggesting a minimum
Doppler factor for the inverse Compton limit of 6.  The simplest
interpretation of the data yields a core size of 0.47\,mas (in one
dimension).  If this is an over-estimate due to scattering in the ISM,
a smaller core size and higher brightness temperature would be
implied.  On the other hand, refractive substructure can have the
effect of enhancing detections on the long baselines to RadioAstron
leading to an over-estimate of the brightness temperature.  Further
RadioAstron observations on baselines less than 6 Earth-diameters are
needed to distinguish between the various possibilities.

ATCA monitoring shows that the source has varied by a factor of 5 in
flux density at 8.4\,GHz, and the {\it Fermi} light-curve suggests
variations in the gamma-ray flux by a factor of at least ten.  A radio
outburst, rising earlier and faster at higher frequencies, followed a
prolonged gamma-ray high state in 2013, with a lag between onsets of
$\sim$9 months, comparable to that seen in other blazars.  Multi-epoch
VLBI observations reveal persistent jet components, with the innermost
component in the first epoch TANAMI image plausibly associated with
the 1996 radio outburst.

The multiwavelength data presented here were combined with other data
to derive an SED, which was fitted using a one-zone synchrotron
model. Despite the covariance between model parameters and the
non-contemporaneous nature of the data, the model~B fit yields a
Doppler factor of $\sim$9, consistent with the lower limit to the
inverse Compton Doppler factor of 6 inferred from the RadioAstron
observation. The equipartition Doppler factor of at least 40 is much
more difficult to reconcile.

\begin{acknowledgements}
The RadioAstron project is led by the Astro Space Center of the
Lebedev Physical Institute of the Russian Academy of Sciences and the
Lavochkin Scientific and Production Association under a contract with
the Russian Federal Space Agency, in collaboration with partner
organizations in Russia and other countries.  The ATCA, Parkes and
Mopra radio telescopes are part of the Australia Telescope National
Facility which is funded by the Commonwealth of Australia for
operation as a National Facility managed by CSIRO.  This study made
use of data collected through the AuScope initiative.  AuScope Ltd is
funded under the National Collaborative Research Infrastructure
Strategy (NCRIS), an Australian Commonwealth Government Programme.
This research was supported by Basic Research Program P-7 of the
Presidium of the Russian Academy of Sciences.
The \textit{Fermi} LAT Collaboration acknowledges generous ongoing
support from a number of agencies and institutes that have supported
both the development and the operation of the LAT as well as
scientific data analysis.  These include the National Aeronautics and
Space Administration and the Department of Energy in the United
States, the Commissariat \`a l'Energie Atomique and the Centre
National de la Recherche Scientifique / Institut National de Physique
Nucl\'eaire et de Physique des Particules in France, the Agenzia
Spaziale Italiana and the Istituto Nazionale di Fisica Nucleare in
Italy, the Ministry of Education, Culture, Sports, Science and
Technology (MEXT), High Energy Accelerator Research Organization (KEK)
and Japan Aerospace Exploration Agency (JAXA) in Japan, and the
K.~A.~Wallenberg Foundation, the Swedish Research Council and the
Swedish National Space Board in Sweden. Additional support for science
analysis during the operations phase is gratefully acknowledged from
the Istituto Nazionale di Astrofisica in Italy and the Centre National
d'\'Etudes Spatiales in France.
This research was funded in part by NASA through {\it Fermi} Guest
Investigator grants NNH09ZDA001N, NNH10ZDA001N, NNH12ZDA001N, and
NNH13ZDA001N-FERMI.  T.H.\ was supported by the Academy of Finland
project number 267324.  Sasha Pushkarev is thanked for helpful
discussions and Sara Buson is thanked for useful comments.  This
research has made use of NASA's Astrophysics Data System, and the
NASA/IPAC Extragalactic Database (NED) which is operated by the Jet
Propulsion Laboratory, California Institute of Technology, under
contract with the National Aeronautics and Space Administration.
\end{acknowledgements}

\end{document}